# The excitation functions of $^{187}$Re(n,2n)$^{186m,g}$Re Reactions

Xiao-Long Huang (黄小龙),[*] Meng-Xiao Kang (康梦霄), Li-Le Liu (刘丽乐),
Ji-Min Wang (王记民), Li-Yang Jiang (江历阳), and Xiong-Jun Chen (陈雄军)
Science and Technology on Nuclear Data Laboratory, China Institute of Atomic Energy,
Beijing 102413,China

A new value for the emission probability of 137.144keV γ-ray of $^{186g}$Re decay are re-recommended to be 9.47±0.03 (%). From this new γ-ray emission probability, the measured cross sections for $^{187}$Re(n,2n)$^{186m}$Re and $^{187}$Re(n,2n)$^{186g}$Re reactions around 14MeV are evaluated, and the total cross section for $^{187}$Re(n,2n)$^{186m+g}$Re reaction at 14.8MeV is recommended to be 2213±116 mb. The UNF code are adopted to calculate the total cross sections for $^{187}$Re(n,2n)$^{186m+g}$Re reaction below 20 MeV fitting to the recommended value 2213±116 mb at 14.8MeV using a set of optimum neutron optical potential parameters which obtained on the relevant experimental data of Re. Then the isomeric cross section ratios for $^{187}$Re(n,2n)$^{186m,g}$Re reaction are calculated using the method of Monte Carlo calculations based on the nuclear statistical theory. Combining these two calculated results, the excitation functions for $^{187}$Re(n,2n)$^{186m}$Re and $^{187}$Re(n,2n)$^{186g}$Re reactions are obtained. The obtained results are in good agreement with the available experimental data, which indicates that present method is useful to deduce the isomeric cross sections for (n,2n) reaction.

## I. INTRODUCTION

The cross sections of neutron induced reactions on rhenium are important for nuclear science and technology. Metal rhenium is one of fusion reactor materials. The cross section for $^{187}$Re(n,2n)$^{186m+g}$Re reaction is an important datum for the safety and environmental evaluation of fusion reactor. On the other hand, $^{186}$Re is an important medically radioisotope. $^{186}$Re is regarded as the best radioisotopes used for radiotherapy and radioimmunotherapy.

Reaction $^{187}$Re(n,2n)$^{186}$Re residual nuclei are both the ground state $^{186g}$Re and isomeric state $^{186m}$Re radioactive isotopes. The half-lives of $^{186g}$Re and $^{186m}$Re are 3.7186 days and 2.0×10$^5$ years. Until now, measurements of these cross sections are very scarce and only provided by several laboratories at 14MeV energy region[1-6]. Among these measurements there existed large discrepancies for $^{187}$Re(n,2n)$^{186g}$Re reaction and large errors for $^{187}$Re(n,2n)$^{186m}$Re reaction. So the theoretical calculation is necessary and interesting.

The aim of this paper is to report reliable excitation functions for $^{187}$Re(n,2n)$^{186m}$Re and $^{187}$Re(n,2n)$^{186g}$Re reactions. For this purpose, nuclear model calculations were done using UNF[7] to obtain the total cross sections for $^{187}$Re(n,2n)$^{186m+g}$Re reaction below 20 MeV; and a Monte Carlo calculations based on the nuclear statistical theory (V-H method) were done to obtain the isomeric cross section ratios for $^{187}$Re(n,2n)$^{186m,g}$Re reaction.

---

[*] huang@ciae.ac.cn

## II. EVALUATION EXPERIMENTAL DATA

All these cross sections for the $^{187}$Re(n,2n)$^{186g}$Re, $^{187}$Re(n,2n)$^{186m}$Re reactions are measured by the activation method. The experimental data are obtained by measuring the activities of $^{186g}$Re product through 137.144keV γ-ray. It's noted that the value adopted in measurements for the emission probability of 137.144keV γ-ray of $^{186g}$Re decay are different ($P_γ$=8.22~8.83%), so the emission probability of 137keV γ ray are first analyzed and evaluated.

### A. Decay data

Measurements of the absolute γ-ray emission probability of 137.144 keV per 100 disintegrations of $^{186g}$Re decay are listed together with evaluated values in Table 1.

**Table 1 Measured and evaluated absolute emission probability of the 137.144 keV γ-ray of $^{186g}$Re decay**

| $P_γ$(137.144keV)/% | References | Comments |
|---|---|---|
| 8.26 ± 0.15 | Goswamy[8] | Not used |
| 9.45 ± 0.16 | Coursey[9] | Same group as Ref. [11], superseded by Ref. [11] |
| 9.39 ± 0.09 | Schonfeld[10] | |
| 9.45 ± 0.08 | Coursey[11] | |
| 9.49 ± 0.03 | Miyahara[12] | |
| 9.35 ± 0.10 | Woods[13] | |
| 9.47 ± 0.03 | | Weighted average, $χ^2$=2.7 |
| 9.47 ± 0.03 | | Adopted value |

A statistical analysis was carried out on the measured values. Our "best" recommended absolute γ-ray emission probabilities are mainly from weighted averages of all measurements except Goswamy[8] and Coursey[9]'s measurements. The evaluated uncertainties are the greatest of the internal or external uncertainties and expanded to cover the most precise input value.

Present evaluation of γ-ray emission probability of 137.144 keV from $^{186g}$Re decay is 9.47±0.03 (%). This new emission probability is used to re-evaluate the measured cross sections for $^{187}$Re(n,2n)$^{186m}$Re and $^{187}$Re(n,2n)$^{186g}$Re reactions.

### B. Analysis of the measured cross sections around 14MeV

Until now, measurements of the cross sections for $^{187}$Re(n,2n)$^{186m}$Re and $^{187}$Re(n,2n)$^{186g}$Re reactions are very scarce and only provided by several laboratories around 14MeV energy region. Among these measurements there existed large discrepancies for $^{187}$Re(n,2n)$^{186g}$Re reaction and large errors for $^{187}$Re(n,2n)$^{186m}$Re reaction. The measured cross sections for $^{187}$Re(n,2n)$^{186m}$Re and $^{187}$Re(n,2n)$^{186g}$Re reactions are summarized in table 2 and 3, respectively.

For $^{187}$Re(n,2n)$^{186m}$Re reaction, there are only two set of measured data at 14.8MeV, and in good agreement with each other within larger errors. It's noted that the measured data from Lu Hanlin et al (EXFOR access # 31406006) which published in J.China Nucl. Phys., 19, (1), 35(1997) is 521 ± 125 mb at 14.77 MeV, and this result was revised by Huang Xiaolong et al. [2]. So the independent measurement from China group is the new report value: 485± 116 mb. The experimental data are obtained by measuring the activities of $^{186g}$Re product through 137keV γ ray, so the γ branching ratio of 137keV γ ray are first corrected and renormalized using present evaluations. The evaluated datum at 14.8MeV is given based on the two corrected values, which also listed in table 2.

**Table 2 Measured and evaluated cross sections for $^{187}$Re(n,2n)$^{186m}$Re reaction**

| Time | Author | En /MeV | σ /mb | Neutron flux | $E_\gamma$ /keV | $P_\gamma$ /% | Corrected cross section* /mb |
|---|---|---|---|---|---|---|---|
| 1997 | Y.Ikeda[1] | 14.80 | 541±189 | $^{93}$Nb(n,2n)$^{92m}$Nb | 137 | 8.22 | 470 |
| 1998 | X.Huang[2] | 14.77 | 485±116 | $^{93}$Nb(n,2n)$^{92m}$Nb | 137 | 8.83 | 452 |
| Present evaluated value at 14.8MeV: 461±106 mb | | | | | | | |

*: Renormalized for decay data only.

For $^{187}$Re(n,2n)$^{186g}$Re reaction, all measurements can be divided into two groups. One is the measurements with lower values, another is T.Fan[3]'s in 1992 with higher values. After corrected by decay data and reference cross sections, such discrepancy still exist. In T.Fan's measurement, the γ branching ratio of 137keV γ-ray is taken as 8.5%. After correcting and renormalizing using present evaluation their data are still higher than those of the first group, but in agreement with Khurana' measurements. The previous data before 1967 are obtained with β⁻ activities of residual products measured. But in T.Fan's work the measurements of the neutron fluence rate are carried out by the absolute(the associated particle counting) and relative(the relative standard cross section) methods, respectively, and their experimental results are in very good consistency.

The decay data used in Karam [5], Druzhinin [6] are not given (or clear) in refs., and selected different monitor reactions compared to T. Fan's measurement. Khurana's measurement[4] is not adopted in evaluation due to the following reasons: (1) the decay data used in his measurements is not clear; (2) reference reaction is $^{56}$Fe(n,p)$^{56}$Mn but the threshold for $^{187}$Re(n,2n)$^{186}$Re reaction is about 7.5MeV. It seems not good to include in present evaluations. Also, Filatenkov's measurement[14] is not adopted in evaluation because there is a discrepancy between his measurement and T. Fan's results. Considering these factors, T.Fan's results seem to be more reasonable and reliable. Present evaluated datum at 14.8MeV is given in table 3, which obtained based on T.Fan's measurements.

**Table 3 Measured and evaluated cross sections for $^{187}$Re(n,2n)$^{186g}$Re reaction**

| Time | Author | En /MeV | σ /mb | Neutron flux | $E_\gamma$ /keV | $I_\gamma$ /% | Corrected cross section /mb |
|---|---|---|---|---|---|---|---|
| 1961 | Khurana[4] | 14.8 | 1675±168 | $^{56}$Fe(n,p)$^{56}$Mn | | | 1775# |
| 1963 | R.A.Karam[5] | 14.1 | 1440±410 | $^{197}$Au(n,2n)$^{196}$Au* | | | 1449# |
| 1967 | A.Druzhinin[6] | 14.8 | 1490±160 | $^{27}$Al(n,α)$^{24}$Na | | | 1490 |
| 1992 | T.Fan[3] | 14.1 | 1966.9±50.3 | $^{93}$Nb(n,2n)$^{92m}$Nb | 137 | 8.5 | 1765.4$ |
| | | 14.6 | 1967.1±44.7 | | | | 1765.5$ |
| | | 14.8 | 1952.0±50.7 | | | | 1752.1$ |
| | | 15.0 | 1903.0±50.3 | | | | 1708.1$ |
| 1997 | A.Filatenkov[14] | 14.85 | 1670±100 | $^{93}$Nb(n,2n)$^{92m}$Nb | 137 | 8.6 | 1517$ |
| Present evaluated value at 14.8MeV : 1752±46 mb | | | | | | | |

*: $T_{1/2}$=5.3d. #: Renormalized for reference cross sections only. $: Renormalized for decay data only.

### C. Evaluated the cross section for $^{187}$Re(n,2n)$^{186m+g}$Re reaction at 14.8MeV

As there are only measurements concentrated on 14MeV energy region, the evaluated cross section for $^{187}$Re(n,2n)$^{186m+g}$Re reaction can be firstly obtained at 14.8MeV. Present evaluation at 14.8MeV is 2213±116 mb, which taken from above evaluated cross section for $^{187}$Re(n,2n)$^{186m}$Re and $^{187}$Re(n,2n)$^{186g}$Re reactions.

## III. MODEL CALCULATION

In order to recommend the excitation functions for $^{187}$Re(n,2n)$^{186g}$Re and $^{187}$Re(n,2n)$^{186m}$Re reactions, the theoretical calculation was performed. The total cross sections for $^{187}$Re(n,2n)$^{186m+g}$Re reaction below 20 MeV are calculated using the UNF code to fit the recommended value 2213±116 mb at 14.8MeV. The isomeric cross section ratios for $^{187}$Re(n,2n)$^{186m,g}$Re reaction are calculated using the method of Monte Carlo calculations based on the nuclear statistical theory. Combining these two calculated results, the excitation functions for $^{187}$Re(n,2n)$^{186m}$Re and $^{187}$Re(n,2n)$^{186g}$Re reactions are obtained.

### A. Optical model and optical potential parameters

The optical model is used to calculated the total, nonelastic, elastic scattering cross sections and elastic scattering angular distribution, and the transmission coefficient of the compound nucleus. The optical potentials considered here are Woods-Saxon form for the real part, Woods-Saxon and derivative Woods-Saxon form for the imaginary parts corresponding to the volume and surface absorption, respectively, and the Thomas form for the spin-orbit part.

The code APOM, by which the best neutron optical potential parameters can be searched automatically with fitting the relevant experimental data of total, nonelastic scattering cross sections and elastic scattering angular distributions, is used to obtain a set of optimum neutron optical potential parameters of $^{187}$Re. Because there are no experimental data of elastic scattering angular distributions for Re, the neutron elastic scattering angular distributions of W, which is neighbors nucleus of Re, is used. A set of optimum neutron optical potential parameters of Re are obtained as follows:

The real part of optical potential,
 $V = 57.4563+0.1556E-0.0224E^2-24(N-Z)/A$
The imaginary part of the surface absorption,
 $Ws = \max\{0.0, 8.0057- 0.4181E-12.0(N-Z)/A\}$
The imaginary part of the volume absorption,
 $Wv = \max\{0.0, 0.06646+0.1891E-0.003214E^2\}$

Where E is the incident neutron energy and $Z$, $N$, $A$ are the number of charge, neutron and mass of the target nucleus respectively.
The spin-orbit couple potential $U$so=6.2.
The radius of the real part, the surface absorption, the volume absorption and the spin-orbit couple potential
 $r$r=1.1870, $r$s=1.26866, $r$v=1.51577, $r$so=1.1870.
The width of the real part, the surface absorption, the volume absorption and the spin-orbit couple potential
 $a$r=0.73796, $a$s=0.4981, $a$v=0.68434, $a$so=0.73796.

Fig. 1 shows comparison of neutron total cross section for natural Re in energy region 0.01 ~ 20 MeV between the theoretical values and experimental data of natural Re. The calculated results are in good agreement with experimental data for energy En >2 MeV. The comparison for calculated results and experimental data of elastic scattering angular distributions for n+ $^{Nat}$Re reaction is given in Fig.2. and the theoretical calculated result is reasonable.

In the view of the analysis shown above, this set of neutron optical potential parameter is used for n+ $^{187}$Re theoretical calculation.

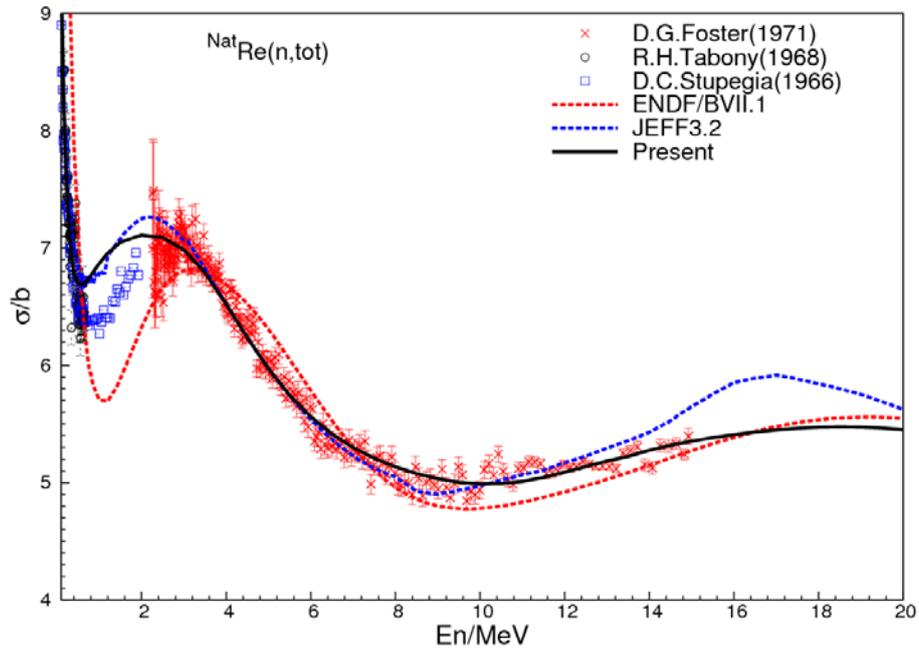

Fig. 1  The total cross section of n+ $^{Nat}$Re reaction

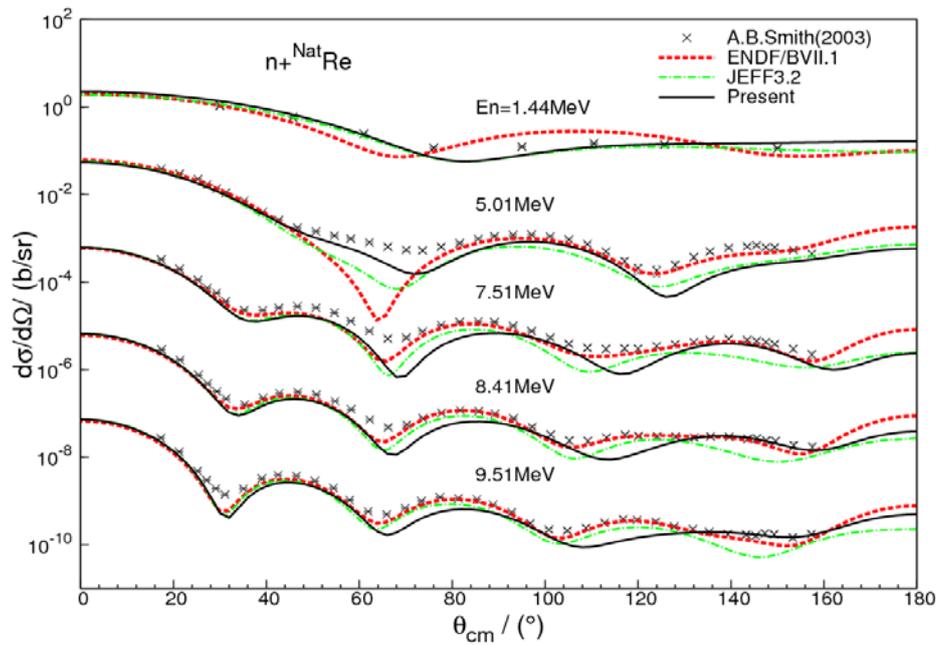

Fig.2  The elastic scattering angular distributions of n+ $^{Nat}$Re reaction

## B. Direct inelastic scattering cross sections

The direct inelastic scattering cross sections to low-lying states are important in uclear data theoretical calculations. The code DWUCK of the distorted wave born approximation was used to calculate the direct reaction cross sections and included as input for the UNF code.

## C. Calculations using UNF

All reaction cross sections, angular distributions, double differential cross sections, and neutron energy spectrum are calculated for n+ $^{187}$Re at incident neutron energies below 20 MeV by the semi-classical theory code UNF with the optical potential parameters shown above, adjusting charged particle optical potential parameters, giant dipole resonance parameters and level density parameters.

The UNF code consists of the optical model, the semi-classical model of multi-step nuclear reaction processes. The semi-classical model of multi-step nuclear reaction processes, in which the discrete level effect in multi-particle emissions as well as the pre-equilibrium phenomenon combining with parity conservation and angular momentum conservation are included, is used to describe the nuclear reaction pre-equilibrium and equilibrium decay processes. This semi-classical model includes both the Hauser-Feshbach theory and the exciton model, and the exact Pauli exclusion effect in the exciton state densities is taken into account. The pick-up mechanism is used to describe the composite particle emission processes. The pre-equilibrium and direct reaction mechanisms of γ emission were also included in this code. The recoil effect was taken into account in the UNF code in order to keep the energy conservation for whole reaction processes.

## D. Calculation the isomeric cross section ratio

Of all the method of calculating the isomeric cross section ratios for (n,2n) reactions, the V-H method[15], which based on the statistical theory, is a frequently used method. From this method, it is necessary to calculate the spin distributions following particle emission. In most cases this method is very simple and rather effective.

The calculation is done within the framework of the spin-dependent statistical theory of nuclear reactions. Usually the important factors which determine the isomeric cross section ratio for (n,2n) reactions are: (1) the spins of the compound system, (2) the angular momentum carried away at each step, (3) the probability of forming states of different spins during each step of the cascade and (4) the spins of the isomeric state and ground state. The compound nuclei of a given excitation energy are formed with a variety of spins. The emitted particle leads to residual nuclei with a variety of spins. When particle emission is not energetically possible, further de-excitation occurs by γ-ray emission and again changes the spin distributions. Ultimately the γ-ray cascade leads to the isomeric state or ground state.

**(1). Initial compound nucleus spin distribution**

The cross section for the formation of a compound nucleus with spin $J_c$ at a bombarding energy E is given by

$$\sigma(J_c,E) = \pi \lambda^2 \sum_{S=|I-s|}^{I+s} \sum_{l=|J_c-S|}^{J_c+S} \frac{2J_c+1}{(2s+1)(2I+1)} T_l(E) \qquad (1)$$

where λ is the de Broglie wavelength of the incoming projectile, s is the spin of the projectile, I is the spin of the target nucleus, S is the entrance channel spin, and $T_l(E)$ is the transmission coefficient of the incident particle of energy E and orbital angular momentum l. Then the normalized initial compound nucleus spin distribution can be written as the following form:

$$P(J_c) = \frac{\sigma(J_c, E)}{\sum_{J_c} \sigma(J_c, E)} \qquad (2)$$

**(2). Spin distribution following neutron emission**

A particular state with spin $J_c$ can decay by particle emission to final states with a variety of spin values, each of which are denoted by $J_f$. The relative probability from an initial state $J_c$ to a final state spin $J_f$ is given by

$$P(J_c \to J_f) \propto \rho(J_f)[D\ (2J_f+1)+(1-D)] \sum_{j=|J_f-s|}^{J_f+s} \sum_{l=|J_c-j|}^{J_c+j} T_l(E_n) \qquad (3)$$

where $T_l(E_n)$ is the transmission coefficient of the emission neutron with angular momentum l and energy $E_n$, D is the contributions from preequilibrium emission which determined by UNF code, and $\rho(J_f)$ is the residual nucleus level density,

$$\rho(J_f) = \frac{2J_f+1}{\sqrt{2\pi^2}\sigma^2} \exp\{-\frac{(J_f+0.5)^2}{2\sigma^2}\} \qquad (4)$$

where $\sigma^2$ is the spin cutoff factor with

$$\sigma^2 = 0.146at\ A^{2/3} = 0.073(1+\sqrt{1+4aU})A^{2/3} \qquad (5)$$

In these formulae, A is the mass number, a is the level density parameter, U=E-Δ, Δ is pairing energy correction. a and Δ are taken from ref. 3.

To sum over all values of $J_c$, one can easily obtain the normalized spin distribution at state $J_f$ following a neutron emission

$$P(J_f) = \sum_{J_c} P(J_c) \cdot \frac{\rho(J_f) \sum_{j=|J_f-s|}^{J_f+s} \sum_{l=|J_c-j|}^{J_c+j} T_l(E_n)}{\sum_{J_F} \rho(J_F) \sum_{j=|J_F-s|}^{J_F+s} \sum_{l=|J_c-j|}^{J_c+j} T_l(E_n)} \qquad (6)$$

The calculation of spin distribution following second neutron emission is repeatedly done by the steps mentioned above, in which case the normalized spin distribution following the emission of the first neutron becomes the initial spin distribution.

## (3) Spin distribution following gamma-ray emission

After the second neutron is emitted, the residual nucleus is de-excited by emitting one or a cascade of gamma rays if energy of the residual nucleus is not enough to emit a neutron. The relative probability from state $J_i$ to state $J_f$ by $\gamma$ emission is assumed to be simply proportional to the density of final state with spin $J_f$. Thus the normalized probability of $J_f$ can be given by the following formula:

$$F(J_f) = \sum_{J_i=|J_f-l|}^{J_f+l} \frac{F(J_i)\rho(J_f)\delta_{J_i,J_f}}{\sum_{J_F=|J_i-l|}^{J_i+l}\rho(J_F)} \qquad (7)$$

where l is the multipolarity of gamma emission and $F(J_i)$ is the initial spin distribution. For the first gamma-ray emission, $F(J_i)=P(J_f)$, where $P(J_f)$ is the spin distribution following the second neutron emission. Considering the pure dipole radiation during gamma de-excitation, the multipolarity l can be taken as 1.

The number of gamma-ray in the gamma-cascade is approximately estimated by

$$N = (l+1)^{-1}\sqrt{aE} = 0.5\sqrt{aE} \qquad (8)$$

and the mean energy of gamma-ray emission at each step can be calculated by

$$E_\gamma = 4(E/a - 5/a^2)^{1/2} \qquad (9)$$

where E is the excitation energy and a is the level density parameter.

The last $\gamma$-ray to be emitted is assumed to lead the excited nucleus to the isomeric state or ground state depending upon which transition has the smaller spin change. Thus of all possible values of $J_f$, one can find a separate spin $I_d$, which makes the state with higher spin than $I_d$ de-excite to the high-spin product and the state with lower than $I_d$ de-excite to low-spin product. Define $I_d$ as the separate spin value

$$I_d = \frac{I_h + I_l}{2} \qquad (10)$$

where $I_h$ is the spin at high spin state and $I_l$ is the spin at low spin state.

If $I_d$ is the possible spin value of final state, the ratio can be expressed

$$r = \frac{\sigma_h}{\sigma_l} = \frac{1 - \sum_{J_f \langle I_d} F(J_f)}{\sum_{J_f \langle I_d} F(J_f)} \qquad (11)$$

and $I_d$ is not the possible spin value of final state, the ratio can be expressed

$$r=\frac{\sigma_h}{\sigma_l}=\frac{1-[\sum_{J_f \langle I_d} F(J_f)+\frac{2J_f+1}{4I_d+2}F(J_f=I_d)]}{\sum_{J_f \langle I_d} F(J_f)+\frac{2J_f+1}{4I_d+2}F(J_f=I_d)} \qquad (12)$$

where $\sigma_h$, $\sigma_l$ represent the cross section of high spin and low spin state, respectively.

So the isomeric cross section ratio R, which isomeric cross section to the sum of isomeric cross section and ground cross section, can be written as follows:

$$R=\frac{r}{r+1} \quad \text{if } I_m>I_g \qquad (13.a)$$

$$R=\frac{1}{r+1} \quad \text{if } I_m<I_g \qquad (13.b)$$

where $I_m$ and $I_g$ represent the spin of isomeric and ground state, respectively.

**(4) Neutron emission energy**

After introducing of the pairing energy and shell energy correction $\delta$, the maximum energy of the outgoing neutron is given by

$$\varepsilon_n = E - S_n - \Delta \qquad (14)$$

where E is the excitation energy and $S_n$ is the neutron separate energy. Defining

$$X_{max}=a^{-1}[(a\varepsilon_n+0.25)^{\frac{1}{2}}-0.5] \qquad (15)$$

the neutron emission energy distribution then becomes

$$f(X)=\frac{X}{X_{max}}\exp\{aX_{max}-[a(\varepsilon_n-X)]^{\frac{1}{2}}\} \qquad (16)$$

where a is the level density parameter. f(X) satisfies the relationship $f(X_{max})=1$.

The actual selection of a value of X involves the consecutive drawing of two random numbers between 0 and 1. The first random number, $\xi_1$, is used to choose a value of X between 0 and $\varepsilon_n$.

$$X=\xi_1\varepsilon_n \qquad (17)$$

The second random number, $\xi_2$, is used to determine whether the value of X is to be accepted or rejected. If $f(X)>\xi_2$, the value of X is accepted, otherwise re-drawing the $\xi_1$ and $\xi_2$.

## IV. THEORETICAL RESULTS AND EVALUATIONS

In order to obtain the excitation functions for $^{187}$Re(n,2n)$^{186g}$Re and $^{187}$Re(n,2n)$^{186m}$Re reactions, the theoretical calculation is performed.

All cross sections of n+$^{187}$Re reactions are calculated by the code UNF firstly. The optical potential parameters, level density and giant dipole resonance parameters are adjusted to make the calculations be consistent with the available measurements around 14MeV, especially with present evaluations for (n,2n) reaction at 14.8MeV. Then the isomeric cross section ratios for $^{187}$Re(n,2n)$^{186m,g}$Re reaction are calculated using the same parameters by the method shown above.

The comparison of different theoretical calculated results from UNF, EMPIRE, TALYS, and present evaluated data at 14.8MeV for $^{187}$Re(n,2n)$^{186}$Re reaction is given in Fig. 3. The calculated results from EMPIRE, TALYS and UNF model calculation are in good agreement each other, and fit present evaluations at 14.8MeV very well. The theoretical calculated result is reasonable.

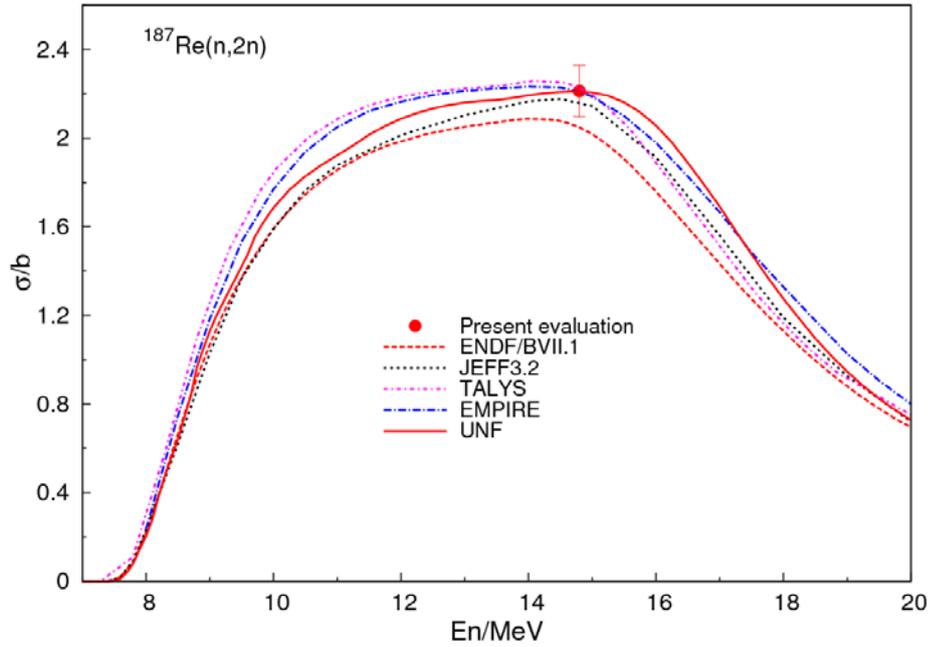

Fig. 3 The cross sections of $^{187}$Re(n,2n)$^{186}$Re reaction

From present evaluation, it's easy to deduce the isomeric cross section ratio R (R=$\sigma_m/\sigma_g$) for $^{187}$Re(n,2n)$^{186}$Re reaction at 14.8MeV. The present calculated results from the method mentioned above are compared with this deduced value, the EMPIRE and TALYS calculated results, which shown in Fig.4. The theoretical calculated isomeric ratio from EMPIRE, TALYS is not in good agreement each other, especially in En>15MeV energy region. The result of V-H method is very close to the evaluated data at 14.8MeV. This indicates that present method is a practical method of calculating the isomeric cross section ratios for (n,2n) reactions. So we conbine the model calculations and the isomer ratio from V-H method to deduce the isomeric cross sections in present work.

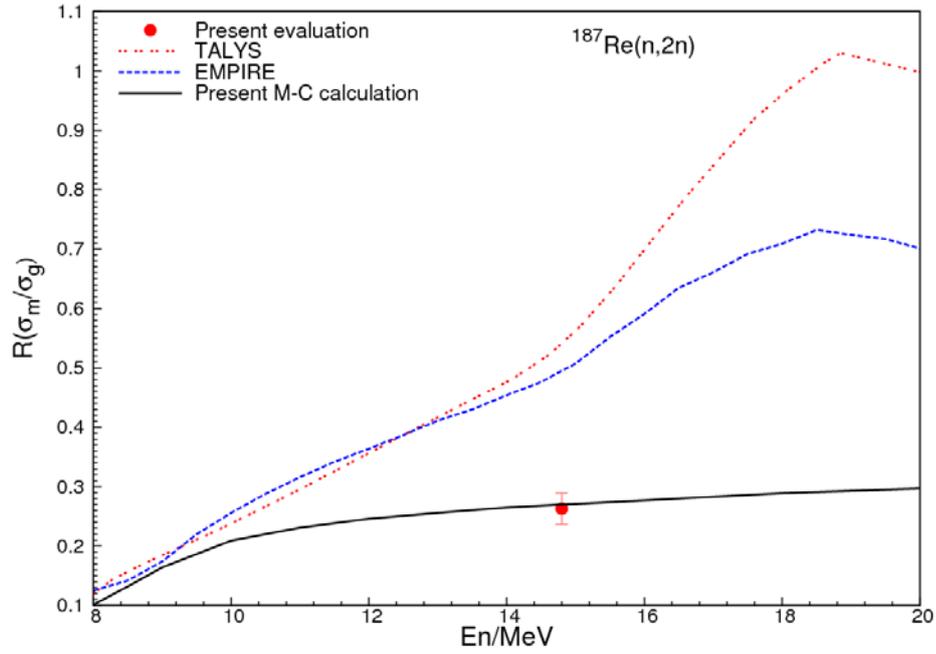

Fig.4 The isomeric cross section ratio for $^{187}$Re(n,2n)$^{186m,g}$Re reaction

Using the calculated total cross section and isomeric cross section ratio for $^{187}$Re(n,2n)$^{186}$Re reaction, the cross sections for $^{187}$Re(n,2n)$^{186m}$Re and $^{187}$Re(n,2n)$^{186g}$Re reactions are obtained, which given in Figs.5 and 6, respectively. It's noted that the obtained cross sections for $^{187}$Re(n,2n)$^{186m}$Re and $^{187}$Re(n,2n)$^{186g}$Re reactions are normalized with present evaluated value 461mb for isomeric, 1752 mb for ground state at 14.8MeV, respectively. And the data plotted in Figs.5 and 6 are the corrected or adjusted results using new decay data or reference cross sections, etc., not the original measured values.

Our recommended cross sections for $^{187}$Re(n,2n)$^{186m}$Re and $^{187}$Re(n,2n)$^{186g}$Re reactions are in good agreement with the available experimental data.

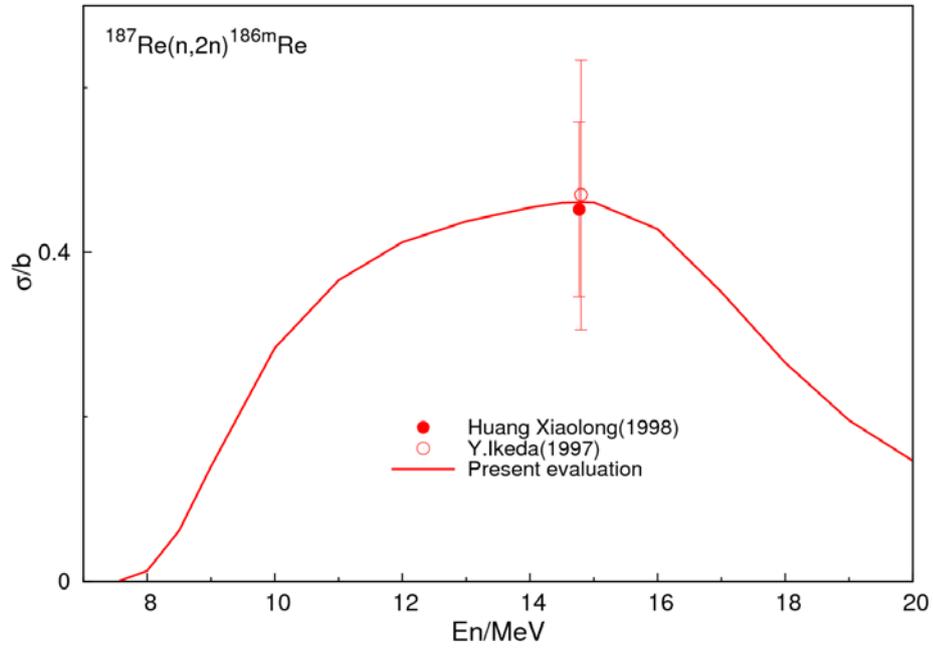

Fig. 5  The cross sections of $^{187}$Re(n,2n)$^{186m}$Re reaction

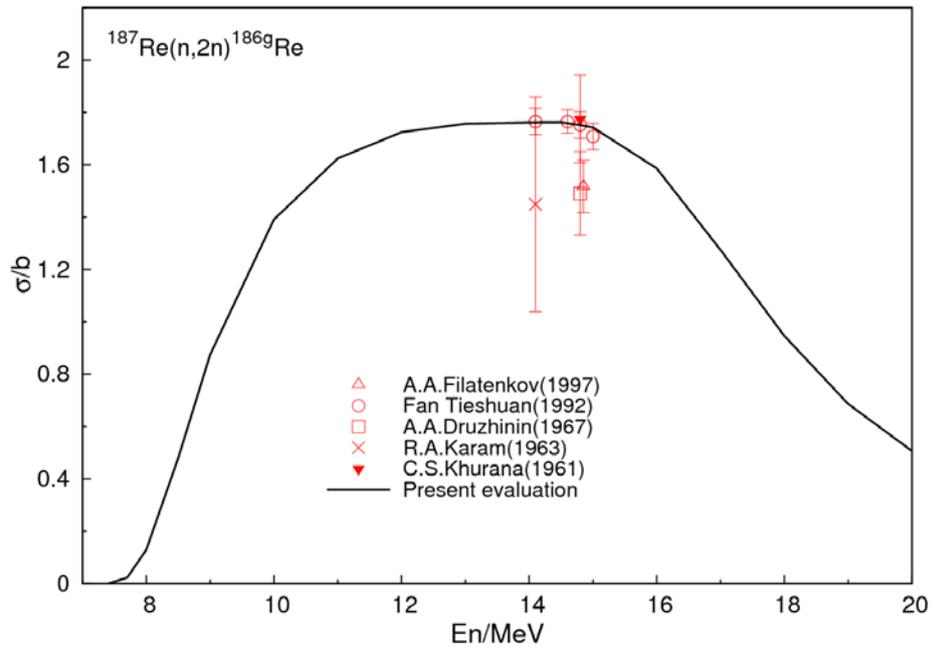

Fig. 6  The cross sections of $^{187}$Re(n,2n)$^{186g}$Re reaction

## V. CONCLUSION

Based on the available experimental data of Re nucleus, a set of optimum optical potential parameters for $0.001 < E_n < 20$ MeV is obtained. With adjusted proton and alpha particle optical potential parameters, level density and giant dipole resonance parameters, all the cross sections of neutron induced reaction on $^{187}$Re are obtained. Because the calculated results for many channels are in pretty agreement with existing experimental data, the predicted cross sections in energy range where there are no any experimental data are reasonable.

The cross sections for $^{187}$Re(n,2n)$^{186m+g}$Re reactions are theoretical calculated based on the available experimental data using UNF code. The isomeric cross section ratios for $^{187}$Re(n,2n)$^{186m,g}$Re reaction are calculated using the V-H method based on the nuclear statistical theory. Their excitation functions for $^{187}$Re(n,2n)$^{186m}$Re and $^{187}$Re(n,2n)$^{186g}$Re reactions are obtained from these theoretical calculations and compared with measurements and other evaluations. From Figs. 1 to 2, one can conclude that present calculations are consistent with the available measurements very well. This means that the parameters used in present calculations are very reliable and reasonable. Of course the calculations should be further checked by the measurements directly in the future because the available experimental data are very scarce and concentrated on 14MeV energy region.

There are several factors which influenced the isomeric ratios. The calculations indicate that the most important factor is the spin cutoff factor. Thus it is very important to select the spin cutoff factor accurately in the calculations.

Now briefly discuss the spin cutoff factor in gamma de-excitation. If allowed it varies with the excitation energy, one can easily find it will be equal to zero. Obviously this result is wrong. So in present work, it is limited a constant during gamma de-excitation. Its value is calculated by eq. (13) with the excitation energy following second neutron emission.